# Surface Dipoles and Work Functions of Alkylthiolates and Fluorinated Alkylthiolates on Au(111)

*Paul C. Rusu and Geert Brocks**

Faculty of Science and Technology and MESA+ Institute for Nanotechnology,

University of Twente, P. O. Box 217, 7500 AE Enschede, The Netherlands..

AUTHOR EMAIL ADDRESS: G.H.L.A.Brocks@tnw.utwente.nl

CORRESPONDING AUTHOR FOOTNOTE   E-mail:   G.H.L.A.Brocks@tnw.utwente.nl; webpage: http://cms.tnw.utwente.nl/

ABSTRACT We study the dipole formation at the surface formed by $-CH_3$ and $-CF_3$ terminated short-chain alkyl-thiolate monolayers on Au(111). In particular, we monitor the change in work function upon chemisorption using density functional theory calculations. We separate the surface dipole into two contributions, resulting from the gold-adsorbate interaction and the intrinsic dipole of the adsorbate layer, respectively. The two contributions turn out to be approximately additive. Adsorbate dipoles are defined by calculating dipole densities of free-standing molecular monolayers. The gold-adsorbate interaction is to a good degree determined by the Au-S bond only. This bond is nearly apolar and its contribution to the surface dipole is relatively small. The surface dipole of the self-assembled monolayer is then dominated by the intrinsic dipole of the thiolate molecules. Alkyl-thiolates increase the work function of Au(111), whereas fluorinated alkyl-thiolates decrease it.

## Introduction

Self-assembled monolayers (SAMs) of organo-thiolate molecules on gold are studied for a wide range of applications, such as supramolecular assembly, biosensors, molecular electronics and microelectronic devices [1-4]. Using organic semiconducting materials as the active components of opto-electronic devices, often the energy barriers for charge injection from metal electrodes into the organic material form a limiting factor for the device performance [5,6]. It has been shown that chemisorption of a SAM on the surface of the metal electrode can alter its work function substantially. By tailoring the SAM's chemical structure this effect can be used advantageously to lower the energy barrier for charge injection and increase the device performance [7,8,9].

The work function change of the surface is directly proportional to the change in the surface electric dipole caused by adsorption of the SAM. Therefore, in order to understand the relation between the work function change and the SAM's chemical structure one has to focus on the dipoles formed in the SAM-metal interface region. One obvious contribution to the surface dipole stems from the permanent dipoles of the molecules within the SAM. It has been demonstrated experimentally that a strong correlation exists between the molecular dipole moments and the work function changes induced by SAMs on gold and silver surfaces [7-10]. The dense packing of molecular dipoles in a SAM, however, causes a sizable depolarizing electric field, which polarizes the molecules such as to effectively reduce their dipole. This effect is often modeled empirically by using an effective dielectric constant for the molecular layer.

A second major contribution to the surface dipole results from the charge reordering associated with the formation of the chemical bonds between the metal surface and the adsorbate molecules. This contribution is foremost determined by the nature of the chemical bonds, but can also be modified by the packing density of the molecules. Thiolate molecules on gold surfaces are among the best studied systems, but it is still debated whether there is a sizable charge transfer between the surface and the molecules upon chemisorption.

In this paper we want to elucidate the role played by the different contributions to the surface dipole of a SAM on gold and study the interplay between them. We calculate the dipole contributions and the work function change from first principles using density functional theory (DFT). In particular, we study alkyl-thiolates on the Au(111) surface, since these are among the best characterized systems, experimentally as well as theoretically [1,11-21]. The common functionals used within DFT are very well suited to describe chemisorption, but lack an accurate description of the van der Waals interactions between the alkyl chains that determine the structure of long-chain alkyl-thiolate SAMs. This inter-chain interaction is relatively unimportant in short-chain alkyl-thiolates and, since we are mainly interested in surface dipole formation, we study the short-chain alkyl-thiolates $CH_3S$ and $CH_3CH_2S$.

The basic building block of the structure of an alkyl-thiolate SAM on Au(111) is well-known. It consists of one thiolate molecule per $\sqrt{3}\times\sqrt{3}\,R30^\circ$ surface unit cell [1,11]. Superstructures of this basic pattern have been reported that contain up to four molecules in the same overall packing density. Experimentally, the positions of the adsorption sites of the thiolate molecules on the surface and the exact structure of the thiolate layers are still hotly debated. Theoretically, the energy differences between several of these structures are very small and are within the error bar of DFT calculations (using common functionals). We examine these structures such as to elucidate as to what extent structural variations lead to a difference in surface dipole.

The sign of the dipole moment of a fluorinated alkyl-thiolate molecule is opposite to that of a non-fluorinated one. Therefore, SAMs of molecules with fluorinated alkyl tails give work function changes that are opposite to those that consist of molecules with normal alkyl tails [7-10]. We analyze the surface dipoles of SAMs containing molecules with $-CF_3$ end groups, in particular $CF_3S$ and $CF_3CH_2S$. The structure of such SAMs is much less well characterized than that of their alkyl counterparts. Long-chain alkyl-thiolates having only $-CF_3$ end groups are believed to have basically the same structure and packing as those with $-CH_3$ end groups, although the $-CF_3$ end groups lead to a larger degree of surface disorder [22]. If long alkyl chains are largely fluorinated, then alkyl-thiolates form a less densely packed SAM [23,24]. A priori it is not clear what SAM structure the molecules $CF_3S$ and $CF_3CH_2S$ would form. Therefore we discuss a couple of possible structures and packings.

## Theoretical section

DFT calculations are performed with the VASP (Vienna *ab initio* simulation package) program [25-27] using the PW91 functional for electronic exchange and correlation [28]. The projector augmented wave (PAW) method is used to represent the electron wave functions [29,30]. For gold atoms, 6s and 5d electrons are treated as valence electrons, for carbon and fluor 2s and 2p, and for sulphur 3s and 3p, respectively. The valence wave functions are expanded in a basis set consisting of plane waves. All plane waves up to a kinetic energy cutoff of 450 eV have been included.

The Au(111) surface is modeled in a supercell containing a slab of typically five or six layers of gold atoms. The SAM is adsorbed on one side of the slab. A vacuum region of 13.1 Å is used, and periodic boundary conditions are applied in all three dimensions. The surface unit cell depends upon the monolayer structure and coverage. Our reference point is a $\sqrt{3} \times \sqrt{3}\, R30^\circ$ surface unit cell, which contains three gold atoms in the surface layer.

The electronic structure is calculated using a uniform **k**-point sampling grid in the surface Brillouin zone (SBZ) and a Methfessel-Paxton broadening of 0.2 eV [31]. A typical **k**-point grid consists of a $8 \times 8$ division of the SBZ of the $\sqrt{3} \times \sqrt{3}\, R30^\circ$ cell. SBZ samplings of other surface cells are chosen such that they have a similar density of grid points. Periodic boundary conditions can lead to spurious interactions between the dipoles of repeated slabs. To avoid such interactions the Neugebauer-Scheffler dipole correction is applied [32]. The electronic structure and the geometry are optimized self-consistently, where typically the positions of the atoms in the SAM and those in the first two layers of the gold slab are allowed to vary. The cell parameter of the Au(111) $1 \times 1$ surface unit cell is fixed at the bulk optimized value of 2.94 Å.

The surface work function $W$ is defined as the minimum energy required to move an electron from the bulk to the vacuum outside the surface and it is given by the expression

$$(1) \quad W = V(\infty) - E_F,$$

where $V(\infty)$ is the electrostatic potential in the vacuum, at a distance where the microscopic potential has reached its asymtotic value; $E_F$ is the Fermi energy of the bulk metal. A self-consistent electronic structure calculation using a plane wave basis set produces the electrostatic potential $V(x,y,z)$ on a grid in real space. Assuming that the surface normal is along the $z$-axis, one can define a plane averaged potential

$$(2) \quad \overline{V}(z) = \frac{1}{A} \iint_{cell} V(x,y,z) dx dy,$$

where $A$ is the area of the surface unit cell. Plotting $\overline{V}(z)$ as function of $z$ is then a convenient way of extracting the value of $V(\infty)$. In practice $\overline{V}(z)$ reaches its asymtotic value already within a distance of 5Å from the surface. An example resulting from a calculation of a SAM of methyl-thiolate $CH_3S$ on Au(111) is shown in Figure 1.

In order to calculate surface work functions according to Eq. (1) one needs an accurate value of the Fermi energy inside the metal. Whereas the value obtained from a slab calculation is quite reasonable, provided a slab of sufficient thickness is used, a better value can be obtained from a separate bulk calculation, following the procedure outlined by Fall *et al.* [33]. Typically DFT calculations give work functions that are within 0.1-0.2 eV of the experimental values, although occasionally somewhat larger deviations are found [34-36].

In order to estimate the convergence of the numbers given in this paper, we perform test calculations in which we vary the **k**-point sampling grid and broadening parameter, the thickness of the slab and of the vacuum region, and the number of layers in which the gold atoms are allowed to relax their positions. From these tests we estimate that the energy differences quoted in this paper are converged to within 1 kJ/mol and the work functions to within 0.05 eV.

## Results and discussion

### *Structures*

In this section we discuss the possible structures of alkyl-thiolate SAMs on Au(111). Our main goal is to study the link between the structure and the work function, which we will discuss in the next section. The earliest experimental (He) diffraction studies established a $\sqrt{3}\times\sqrt{3}\,R30^\circ$ structure for alkyl-thiolate SAMs on Au(111) with one molecule per surface unit cell [37,38], see Figure 2. Somewhat later a $c(4\times2)$ superstructure was found, which contains four thiolate molecules per surface unit cell in the same packing density as the simpler $\sqrt{3}\times\sqrt{3}\,R30^\circ$ structure [39], see Figure 3. From infrared data it was concluded that there are two different orientations in the alkyl chains [40] and from grazing incidence X-ray diffraction data, a model was proposed for the superstructure based upon thiolate dimers [41]. Evidence against the dimer model was presented by scanning tunneling microscopy [42] and by electron spectroscopy experiments [43]; in the latter, thiolate dimers were found only at temperatures above 375K. In recent He and X-ray diffraction experiments it was concluded that in the $c(4\times2)$ superstructure alkyl-thiolate molecules adsorb as monomers on the Au(111) surface [44,45]. Experimentally the $c(4\times2)$ and $\sqrt{3}\times\sqrt{3}\,R30^\circ$ structures seem to be close in energy; in scanning tunneling microscopy experiments domains of both structures have been shown to coexist [22]. Moreover, larger and more complex superstructures such as $(3\times4)$ could also be close in energy [46].

Regarding the exact binding sites of thiolate molecules, diffraction and time of flight scattering studies emphasize the hollow sites on the Au(111) surface, where the sulphur atoms are threefold coordinated by Au atoms of the substrate [45,47]. From recent photoelectron diffraction data and X-ray standing wave analysis it was concluded, however, that thiolate molecules favor the on-top adsorption sites, where a sulphur atom is positioned on top of a single Au atom of the substrate [48,49].

The adsorption of Methyl-thiolate $CH_3S$ on Au(111) has been studied intensively by first-principles calculations in recent years. Most of these calculations consider the basic $\sqrt{3}\times\sqrt{3}\,R30^\circ$ structure, [12-18] and a number of them have addressed the $c(4\times2)$ superstructure [14-20]. Earlier calculations give the threefold hollow sites on the Au(111) surface as the most stable sites for adsorption of the thiolate molecules [12,13,18], but more accurate recent calculations distinctly prefer the twofold bridge sites [14-17,20]. The S-atom of the adsorbate molecule is bonded to two Au-atoms of the surface, see Figure 2. The on-top adsorption site is clearly unfavorable; in most calculations it not even represents a metastable structure, but a maximum on the energy surface. To achieve converged computational results it has become evident from these calculations that the number of Au layers representing the substrate has to be sufficiently large, and that relaxation of the surface atoms is significant. Moreover, it is important to have a sufficiently dense Brillouin zone sampling.

Different density functionals give somewhat different values for the adsorption energy of alkyl thiolates on Au(111), but they favor the same order in preferential binding sites, i.e. the bridge site is much more stable than the hollow site, which is more stable than the on-top site. In addition, calculations on small clusters indicate that DFT and Hartree-Fock (plus many-body perturbation or coupled cluster corrections) give essentially the same stable structures [50]. Calculations on the $c(4\times2)$ superstructure clearly favor adsorption of thiolate molecules as monomers instead of dimers, in agreement with recent experimental results. Because of the strong preference for the bridge adsorption site, most $c(4\times2)$ superstructures that have been proposed from calculations are based upon molecules adsorbed at different bridge sites [14,16,17], see Figure 3. However, the calculated total energy differences between such $c(4\times2)$ and the $\sqrt{3}\times\sqrt{3}\,R30^o$ structures are $\leq 5$ kJ/mol. Such energy differences are too small to be reproduced accurately by common density functionals.

In view of these computational and experimental results we consider only structures in which the alkyl-thiolates are adsorbed as monomers on the Au(111) surface. First we focus upon CH$_3$S on Au(111) in the $\sqrt{3}\times\sqrt{3}\,R30^o$ structure as shown in Figure 2(a),(b). As in previous calculations we find that the bridge site is more stable than the hollow site and that the on-top position is unstable. The relative energies associated with these adsorption sites are given in Table I, in the columns marked by "bridge(s)", "fcc hollow", and "on-top". Table I also presents some structural data. Energies and structures are in fair agreement with the results obtained in previous calculations [14,16,17]. The spread in the results obtained in different calculations reflect the use of different density functionals, as well as slightly different computational parameters.

The $\sqrt{3}\times\sqrt{3}$ R30° structure of $CH_3S$ adsorbed at the bridge site on Au(111) as shown in Figure 2(a), (b), has mirror and glide plane symmetry (this structure has the two-dimensional space group Cm). Rotating the $CH_3$-group around the CS bond breaks the mirror and glide plane symmetry. Breaking the symmetry and optimizing the geometry results in a structure shown in Figure 2(c),(d). The structural data of $CH_3S$ at the bridge site in this 'broken symmetry' structure, bridge(bs), are also given in Table I. This structure is quite similar to the symmetric bridge(s) structure. The most significant change in the bridge(bs) structure, besides the $CH_3$ rotation already mentioned, are that the two Au-S bonds have become slightly inequivalent; moreover, the azimuthal angle $\phi$ of the CS bond has changed. The calculated total energies of the bridge(s) and bridge(bs) structures are very close, the latter being actually 1.7 kJ/mol lower in energy. However, this is within the error bar associated with the PW91 functional we used.

Since the bridge site is much more stable than other adsorption sites, it is reasonable to base a $c(4\times2)$ superstructure entirely upon molecules adsorbed at bridge sites. Figure 3 shows the $c(4\times2)$ superstructure proposed by Vargas *et al.* [14], which contains four molecules per cell. The inequivalent molecules have a similar tilt angle of the CS bond with respect to the surface normal, but differ by ~60° in the azimuthal angle $\phi$ of that bond. We optimized this structure using the same computational parameters as for the $\sqrt{3}\times\sqrt{3}$ R30° unit cell calculations (in particular the **k**-point grid for the Brillouin zone integration). The results are listed in Table I. The local geometries of all molecules in the $c(4\times2)$ structure are quite similar, so we only give the average geometric parameters. The structural parameters are actually similar to those of the $\sqrt{3}\times\sqrt{3}$ R30° bridge(s) structure. The largest difference is in the C-S-normal angle, where the molecules in the $\sqrt{3}\times\sqrt{3}$ R30° structure are tilted somewhat more upright. The total energy of the $c(4\times2)$ structure is only 1.0 kJ/mol per molecule higher than that of the $\sqrt{3} \times \sqrt{3}$ R30° bridge(s) structure, so again it is hardly possible to distinguish between these two structures energetically on the DFT level.

In conclusion, from a computational point of view there are several structures of a CH$_3$S SAM on Au(111) that are very close in energy and are based upon adsorption of CH$_3$S molecules on bridge sites. They differ in the azimuthal angle $\phi$ of the CS bond and/or the rotation angle of the CH$_3$-group around the CS bond. The spread in the tilt angle C-S-Au of the CS bond with respect to the surface is smaller. The possible structures of an ethyl-thiolate (CH$_3$CH$_2$S) SAM on Au(111) very much resemble those of a methyl-thiolate SAM. Table II gives the energy and geometry of CH$_3$CH$_2$S in the $\sqrt{3} \times \sqrt{3}$ R30° structure with the molecule adsorbed on a bridge site in the symmetric (bridge(s)), as well as in the broken symmetry (bridge(bs)) structure. A comparison to the data shown in Table I shows that indeed the geometries of the adsorbed alkyl-thiolates are very similar.

The structure of (partially) fluorinated alkyl-thiolates on gold is much less well-established than that of their alkyl counterparts, both experimentally and theoretically. SAMs of long-chain alkyl-thiolates with fluorinated (−CF$_3$) end groups show a large degree of surface disorder, although their basic structure remains similar to that of SAMs of alkyl-thiolates with −CH$_3$ end groups [22]. Our calculations on trifluoromethyl-thiolate (CF$_3$S) on Au(111) show that within a $\sqrt{3} \times \sqrt{3}$ packing of molecules similar variations in structure are possible as for CH$_3$S on Au(111), with similar small energy differences between these structures.

Thiolates with a longer fluorinated alkyl chain form SAMs with a less denser packing, because of the bulkier fluorinated alkyl tails. Whereas in a $\sqrt{3} \times \sqrt{3}$ packing the nearest neighbor distance between the sulphur atoms on the surface is 5.0 Å, see Figure 2, in fluorinated alkyl-thiolates this distance is 5.8 Å, leading to a 30% less dense surface packing [11,23,24]. This distance is twice the nearest neighbor distance between the gold atoms in the surface and a $p(2 \times 2)$ structure has been proposed for fluorinated alkyl-thiolate SAMs on Au(111) [23]. More complex superstructures, such as a $c(7 \times 7)$, or even an incommensurate structure have also been proposed [24,51]. For the short-chain fluorinated molecules we are considering, i.e. CF$_3$S and CF$_3$CH$_2$S, it is not a priori clear whether the dense $\sqrt{3}\times\sqrt{3}$ or the less dense $2\times2$ packing is favored. Therefore we have also optimized the geometries of these molecules in a $p(2\times2)$ unit cell. An example of an optimized structure is shown in Figure 4. Also in this structure the (displaced) bridge site is the most favorable site for adsorption. The local geometry of the adsorbed molecules is in fact quite similar to that in the $\sqrt{3}\times\sqrt{3}$ packing, as is demonstrated by Table III. We will consider both possible packings in discussing the work functions.

### *Work functions and surface dipoles*

We now focus upon the distribution of dipole moments at the SAM-gold surface. A sensitive technique for characterizing a surface dipole, experimentally or theoretically, is to determine the work function of the surface. For the clean Au(111) surface we calculate a work function of 5.25 eV. Reported experimental values are 5.26 eV [52], and 5.35 eV [21] and previously reported calculated values are 5.23 eV [53], 5.27 eV [54], 5.31 eV [55], and 5.35 eV [21]. All these values are within the experimental and computational error bars.

The calculated work functions of the SAM-Au(111) surfaces in the different geometries are given in Table I-III. It can be observed that structures whose total energies are close, also have similar work functions. For instance, the total energies of the bridge(s) and the bridge(bs) $\sqrt{3}\times\sqrt{3}$ R30° structures of CH$_3$S on Au(111), as well as that of the $c(4\times2)$ structure, are within 2 kJ/mol of one another, see Table I. The work functions of the bridge(s) and bridge(bs) structures differ by 0.04 eV, which is close to the (convergence) error bar of the DFT calculations. The work function of the $c(4\times2)$ structure is ~0.2 eV higher. As compared to the clean Au(111) surface, the work function is shifted to a substantially lower value for these three structures.

The latter involve a similar bonding of molecules at bridge sites and also the local geometry of the adsorbed CH$_3$S molecules is very similar. Comparing the geometries of the bridge and the $c(4\times2)$ structures in Table I, one observes that the bond lengths are similar, as well as the Au-S-Au bond angle. The C-S-normal tilt angle in the $c(4\times2)$ structure is 11-12$^0$ higher than in the $\sqrt{3}\times\sqrt{3}$ R30° structures, which means that in the latter the thiolate molecules are standing somewhat more upright. As a result the component of the molecular dipole moment along the surface normal is somewhat larger in the $\sqrt{3}\times\sqrt{3}$ R30° structures. The work function shift, i.e. the difference between the work functions of the SAM covered surface and that of the clean Au(111) surface, is sensitive to this normal component; see also the analysis below. The fact that the work function shift of both $\sqrt{3}\times\sqrt{3}$ R30° structures is ~0.2 eV larger that of the $c(4\times2)$ structure can be attributed to a larger normal component of the molecular dipoles.

The azimuthal angle $\phi$ and the rotation angle around the CS bond are substantially different in the three structures, see Figure 2 and Figure 3. However, these angles do not affect the normal molecular dipole component and hence they do not affect the work function. The same geometrical arguments also hold for the other molecules. As can be observed from Table II and Table III, one obtains similar work functions in the bridge(s) and bridge(bs) $\sqrt{3}\times\sqrt{3}$ R30° structures for all molecules. In conclusion, for the SAMs we have studied, the structures that have nearly the same total energy also have a very similar local geometry and hence a very similar work function.

Comparing in Table I the work functions of $\sqrt{3}\times\sqrt{3}$ R30° structures that correspond to adsorption of $CH_3S$ at different sites, one observes that the fcc hollow site leads to a work function that is ~0.4 eV than that of the bridge site, whereas that of the on-top site is ~0.9 eV higher. Following the arguments presented above, this can be partly understood from the difference in the C-S-normal tilt angle between these structures. A small (large) tilt angle gives a large (small) normal molecular dipole component and a large (small) shift in work function with respect to the clean Au(111) surface. In addition, the bonding between the molecules and the surface, which is different for the different adsorption sites, contributes to the work function shift. We suggest that work function measurements might contribute to settle the issue of the adsorption site of $CH_3S$ on Au(111) from an experimental point of view.

From Table III one can also observe that the work functions of the fluorinated alkyl-thiolate SAMs on Au(111) do not depend strongly upon the packing density of the molecules. Both the $\sqrt{3}\times\sqrt{3}$ R30° and the $p(2\times 2)$ structures, whose packing density differs by 33%, have a similar work function. This is slightly surprising since one expects the shift in the work function upon adsorption of the SAM to depend upon the packing density of the molecules. One the one hand, one would expect a higher packing density to result in a larger work function shift because of a higher density of molecular dipoles. On the other hand, increasing the packing density of the molecular dipoles also enlarges the depolarizing field in the SAM, which acts to decrease the dipoles and therefore the work function shift. Apparently in the range of packing densities that are relevant to the $\sqrt{3}\times\sqrt{3}\,R30^o$ and $p(2\times 2)$ structures, these two effects tend to cancel one another.

## *Analysis*

We will use the bridge(s) $\sqrt{3}\times\sqrt{3}\,R30^o$ structure to analyze the work functions and begin by noticing that the latter fall into two groups. Adsorption of the two alkyl-thiolates $CH_3S$ and $CH_3CH_2S$ gives one work function, compare Table I and Table II, whereas adsorption of fluorinated $CF_3S$ and $CF_3CH_2S$ molecules gives another value for the work function, see Table III. It makes sense to use the work function of the clean Au(111) surface, 5.25 eV, as a reference point and analyze changes of the work functions upon adsorption of the molecules. We define the change in work function caused by the SAM with respect to the clean Au(111) surface as $\Delta W = W_{SAM} - W_{metal}$. The results are collected in Table IV, which shows that SAMs of $CH_3S$ and $CH_3CH_2S$ lower the work function by 1.3-1.4 eV, whereas SAMs of $CF_3S$ and $CF_3CH_2S$ increase the work function by 0.7-1.0 eV.

These work function changes $\Delta W$ can be interpreted in terms of changes in the surface dipole $\Delta\mu$ due to adsorption of the SAM. Simple electrostatics gives the relation [56]

(3) $$\Delta W = \frac{e\Delta\mu}{\varepsilon_0 A},$$

where $A$ is the surface area taken up by one molecule and $\Delta\mu$ is the change in surface dipole that occurs upon adsorption of the SAM, normalized per molecule. Note that $\Delta\mu$ corresponds to the component of the dipole moment directed along the surface normal, since it is only this component that affects the work function. The values of $\Delta\mu$ calculated according to Eq. ( 3 ) are also given in Table IV. The sign of $\Delta\mu$ is such that for CH$_3$S and CH$_3$CH$_2$S the dipoles point from the surface into the gold crystal, whereas for CF$_3$S and CF$_3$CH$_2$S they point from the surface into the vacuum. Intuitively one would like to interpret $\Delta\mu$ in terms of molecular dipole moments and indeed the size of $\Delta\mu$ is of the order of a molecular dipole moment [7-10]. One should bear in mind however that upon adsorption on a metal surface, even molecules that have a zero dipole moment can alter the surface dipole considerably [57,58].

Therefore we divide $\Delta\mu$ into a part that results from the molecules only and a part that results from the formation of chemical bonds between the molecules and the surface

$$( 4 ) \quad \Delta\mu = \mu_{SAM} + \mu_{chem} .$$

We *define* $\mu_{SAM}$ as the dipole moment along the surface normal of a molecule *embedded in a free-standing* SAM, i.e. without the presence of the metal substrate. This definition takes care of the geometry of the molecule in the SAM. In addition, this definition automatically incorporates the effect of the depolarizing electric field that is caused by the molecular dipoles surrounding each molecule in the SAM. The effect of this depolarizing field is sometimes introduced phenomenologically as an effective dielectric constant in the molecular layer [7-10]. Using a computational technique in which periodic boundary conditions are applied, there is no need for a phenomenological dielectric constant, since the calculation is done automatically on a full monolayer.

One places the molecule in a unit cell in the correct SAM geometry and performs a self-consistent DFT calculation while keeping the geometry fixed. Electrostatics then relates $\mu_{SAM}$ to the step in the electrostatic potential as [56]

$$( 5 ) \quad V_{SAM}(\infty) - V_{SAM}(-\infty) = \frac{e\mu_{SAM}}{\varepsilon_0 A} ,$$

where $V_{SAM}(\infty)$, $V_{SAM}(-\infty)$ are the asymtotic electrostatic potentials on both sides of the SAM. These are easily obtained from the calculation, since the potential reaches its asymtotic values within a distance of few Å from the molecular layer, as is illustrated in Figure 5. The calculated $\mu_{SAM}$ are listed in Table IV.

Having calculated $\Delta\mu$ and $\mu_{SAM}$ from Eqs. (3) and (5), respectively, we then define $\mu_{chem}$ by Eq. (4) and interpret it as the change in the surface dipole due to chemisorption of the molecule on the metal surface. It reflects the charge reordering in the molecule and in the metal surface that takes place upon formation of a chemical bond. The calculated $\mu_{chem}$ are given in Table IV for the four molecules. Two main conclusions can be drawn from these numbers. Firstly, $\mu_{chem}$ shows relatively little variation within this range of molecules. Apparently the charge reordering is mainly confined to the chemical bond formed at the surface, which for all four molecules is the gold-sulphur bond, and it does not depend much on the (fluorinated) alkyl tail of the molecules. Secondly, the absolute value of $\mu_{chem}$ is very small as compared to the absolute value of $\mu_{mol}$. This means that the charge reordering in the gold-sulphur bond is such that it does not give rise to a substantial dipole moment. In other words, the gold-sulphur bond is nearly apolar. This result agrees with that obtained in a previous calculation on $CH_3S$ on Au(111) [21].

Kelvin probe measurements of the work functions of SAMs of the long chain thiolates $C_{16}H_{33}S$ and $C_8F_{17}C_2H_4S$ on Au(111) have been reported by de Boer *et al.* [9]. Work function changes upon SAM adsorption on Au(111) have been deduced from photoemission measurements by De Renzi *et al.* [21] for $CH_3S$, and by Alloway *et al.* [10] for alkyl-thiolates ranging from $C_3H_7S$ to $C_{18}H_{37}S$ and for fluorinated alkyl-thiolates ranging from $CF_3C_{12}H_{24}S$ to $CF_3C_{15}H_{30}S$, and from $CF_3C_{15}H_{30}S$ to $C_{10}F_{21}C_6H_{12}S$. The trends in these measurements are clear; alkyl-thiolates lead to a substantial lowering of the work function relative to the clean gold surface, whereas (partially) fluorinated alkyl-thiolates give an increase of the work function. For alkyl-thiolates photoemission gives a $\Delta W$ of $-1.2$ eV for $CH_3S$ [21], and a $\Delta W$ ranging from $-1.0$ eV for $C_3H_7S$ to $-1.4$ eV for $C_{16}H_{33}S$ [10]. The Kelvin probe gives a $\Delta W$ of $-0.8$ eV for $C_{16}H_{33}S$ [9].

Our calculated value for CH$_3$S is in good agreement with the experimental value, see Table IV. Also the value we find for CH$_3$CH$_2$S is within the range determined by the photoemission experiments. The work function shows only little variation for alkyl-thiolates because both $\mu_{SAM}$ and $\mu_{chem}$ only weakly depend upon the length of the alkyl tail. The dipole $\mu_{chem}$ resulting from chemisorption of the alkyl-thiolate is mainly determined by the gold-sulphur bond. Both the dipole moment of an alkyl-thiolate molecule and the orientation of the molecule in the SAM do not vary strongly with the size of the alkyl tail. Therefore, $\mu_{SAM}$ only weakly depends upon the size of the alkyl group. The Kelvin probe measurement gives a somewhat lower value for $\Delta W$ than the photoemission experiments. This could be due to a number of reasons. The experiments are not performed in UHV, which might introduce impurities. Under ambient conditions the work function of a clean gold surface becomes 4.9 eV instead of the 5.3 eV obtained under UHV conditions [9]. Incorporating this difference would give a $\Delta W$ of $-1.2$ eV, which would bring it close to the photoemission and to the calculated results.

A direct comparison to the experimental results on fluorinated alkyl thiolates is more difficult. The dipoles $\mu_{SAM}$ of fluorinated alkyl-thiolates vary more widely than those of unsubstituted alkyl-thiolates. The molecular dipoles depend on which and how many of the hydrogens on the alkyl tail are substituted by fluor. The substitution also affects the structure of the SAM, and hence the depolarizing field. Moreover, SAMs of fluorinated alkyl-thiolates tend to show more intrinsic disorder than their unsubstituted counterparts. Photoemission gives a maximum $\Delta W$ of $+0.5$ eV for C$_{10}$F$_{21}$C$_6$H$_{12}$S and the Kelvin probe gives $+0.6$ eV for C$_8$F$_{17}$C$_2$H$_4$S. The calculations on short-chain fluorinated alkyl-thiolates give somewhat larger $\Delta W$'s as can be observed from Table IV.

## Conclusions

We have studied the surface dipoles and work functions of SAMs of CH$_3$S, CH$_3$CH$_2$S, CF$_3$S and CF$_3$CH$_2$S on the Au(111) by means of density functional theory calculations. Several structures exist that have almost the same total energy, but also give very similar work functions. By performing separate calculations on the free-standing SAMs we can calculate the dipole moments of the molecules as they are embedded in the SAMs. This allows us to define dipole moments that result from chemisorption of the molecules on the Au(111) surface. The latter are almost independent of the molecule, indicating that they mainly result from the gold-sulphur bond. Moreover, the gold-sulphur bond turns out to be apolar and give only a small contribution to the surface dipoles. The main contributions to the latter stems from the molecular dipole moments. The direction of these is such that adsorption of $CH_3S$ and $CH_3CH_2S$ lowers the work function as compared to the clean Au(111) surface, whereas adsorption of $CF_3S$ and $CF_3CH_2S$ increases the work function.

ACKNOWLEDGMENT  We thank B. de Boer for introducing us to this topic and for showing us his experimental data prior to publication.  P. J. Kelly is acknowledged for useful discussions and G. de Wijs and M. Uyttewaal are thanked for help with the VASP program. This work is supported by the "Prioriteits Programma Materialenonderzoek" (PPM), and is part of the research program of the "Stichting voor Fundamenteel Onderzoek der Materie" (FOM), financially supported by the "Nederlandse Organisatie voor Wetenschappelijk Onderzoek" (NWO). The use of supercomputer facilities was sponsored by the "Stichting Nationale Computer Faciliteiten" (NCF), financially supported by NWO.

FIGURE CAPTIONS

**Figure 1.** Plane averaged electrostatic potential $\overline{V}(z)$ of a slab comprising six layers of gold atoms and one layer of methyl-thiolate $CH_3S$. The $z$-axis is along the 111 direction. Indicated are the Fermi energy $E_F$, the work function $W_{SAM}$ of the SAM and of the clean metal $W_{metal}$.

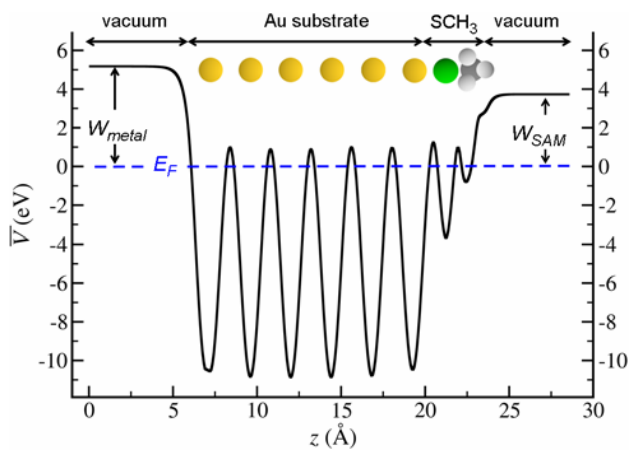

**Figure 2.** The $\sqrt{3}\times\sqrt{3}$ $R30°$ structure of the $CH_3S$ SAM on Au(111) with the molecules adsorbed at bridge sites; a), b) top and side view of the bridge (s, symmetric) structure; c), d) top and side view of the bridge (bs, broken symmetry) structure.

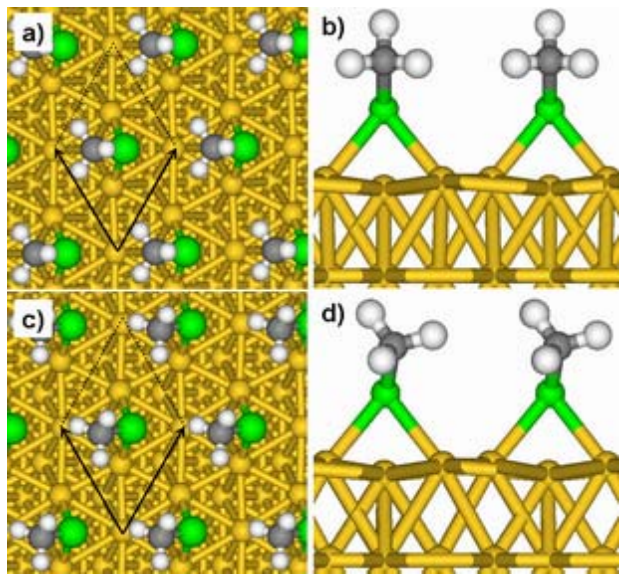

**Figure 3.** Top view of the $c(4\times 2)$ structure of the CH$_3$S SAM on Au(111), which contains 4 molecules per surface unit cell [14].

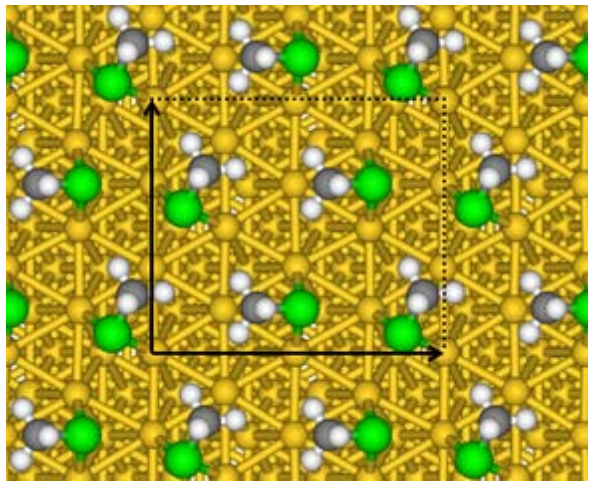

**Figure 4**. Top view of the $p(2\times 2)$ structure of the CF$_3$S SAM on Au(111).

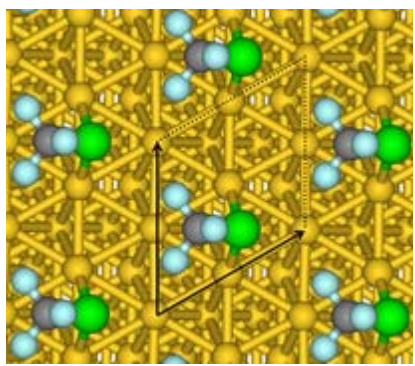

**Figure 5.** Plane averaged electrostatic potential $\overline{V}(z)$ of a free-standing monolayer of methyl-thiolate CH$_3$S. Indicated are the asymtotic electrostatic potentials $V(\infty)$, $V(-\infty)$ on both sides of the layer.

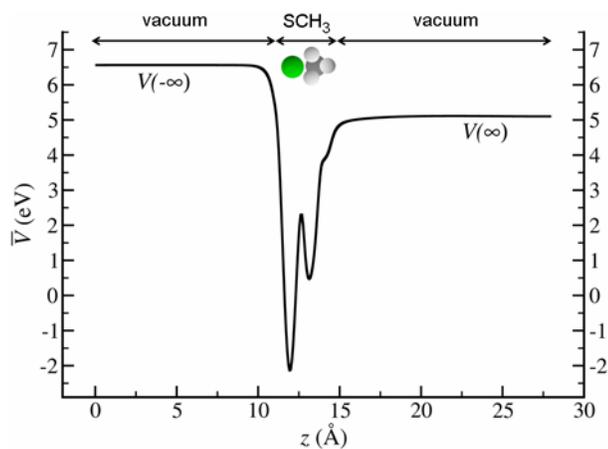

**Table I.** Total energies, bond lengths, bond angles, and work functions of SAMs of CH$_3$S on Au(111). The columns indicate the possible adsorption sites in the $\sqrt{3}\times\sqrt{3}\,R30°$ or $c(4\times 2)$ structures. For geometries in the $c(4\times 2)$ structure only averages over the four molecules are given.

|  | $\sqrt{3}\times\sqrt{3}\,R30°$ | | | | $c(4\times 2)$ |
|---|---|---|---|---|---|
|  | bridge(s) | bridge(bs) | fcc hollow | on-top |  |
| Energy (kJ/mol) | 0.0 | −1.7 | 25.3 | 34.9 | 1.0 |
| Au-S (Å) | 2.50 /2.50 | 2.52 /2.49 | 2.63/2.60 /2.50 | 2.39 | 2.52 /2.50 |
| S-C (Å) | 1.84 | 1.84 | 1.85 | 1.82 | 1.83 |
| Au-S-Au (°) | 77.0 | 76.5 | 74.6/76.7 /76.3 | - | 73.2 |
| C-S-normal (°) | 45.3 | 46.3 | 14.4 | 64.6 | 57.2 |
| $\phi$ (°) | 30.0 | 38.4 | - | - | 31.8 |
| W (eV) | 3.81 | 3.85 | 3.39 | 4.73 | 4.04 |

**Table II.** Total energies, bond lengths, bond angles, and work functions of SAMs of CH$_3$CH$_2$S on Au(111) in the $\sqrt{3}\times\sqrt{3}\,R30°$ structure. The columns bridge(s) and bridge(bs) indicate the symmetric and 'broken symmetry' bridge sites, respectively.

|  | bridge(s) | bridge(bs) |  | bridge(s) | bridge(bs) |
|---|---|---|---|---|---|
| Energy (kJ/mol) | 0.0 | −1.0 | Au-S-Au (°) | 74.5 | 76.8 |
| Au-S (Å) | 2.54 | 2.52/2.48 | C-S-normal (°) | 54.6 | 52.4 |
| S-C (Å) | 1.85 | 1.85 | $\phi$ (°) | 30.0 | 26.2 |
| C-C (Å) | 1.53 | 1.52 | W (eV) | 3.93 | 3.83 |

**Table III.** Bond lengths, bond angles, and work functions of SAMs of $CF_3S$ and $CF_3CH_2S$ on Au(111) in the $\sqrt{3}\times\sqrt{3}\,R30°$ and $p(2\times2)$ structures. The columns bridge(s) and bridge(bs) indicate the symmetric and 'broken symmetry' bridge sites in the $\sqrt{3}\times\sqrt{3}\,R30°$ structure, respectively.

|  | $CF_3S$ | | | $CF_3CH_2S$ | | |
|---|---|---|---|---|---|---|
|  | bridge(s) | bridge(bs) | $p(2\times2)$ | bridge(s) | bridge(bs) | $p(2\times2)$ |
| Au-S (Å) | 2.50 | 2.51/2.47 | 2.52/2.53 | 2.52 | 2.54/2.50 | 2.52/2.50 |
| S-C (Å) | 1.85 | 1.86 | 1.85 | 1.85 | 1.84 | 1.85 |
| C-F (Å) | 1.36 | 1.36 | 1.36 | 1.37 | 1.36 | 1.37 |
| Au-S-Au (°) | 77.0 | 78.4 | 73.4 | 76.4 | 75.6 | 74.6 |
| C-S-normal (°) | 45.7 | 37.8 | 49.8 | 43.3 | 52.0 | 50.6 |
| $\phi$ (°) | 30.0 | 21.1 | 29.6 | 30.0 | 24.2 | 29.5 |
| W (eV) | 5.97 | 5.98 | 6.00 | 6.27 | 6.35 | 6.21 |

**Table IV.** Work functions of SAMs on Au(111) in eV ion the bridge(s) $\sqrt{3}\times\sqrt{3}\,R30°$ structure; absolute values $(W)$ and relative to clean Au(111) $(\Delta W)$. The surface dipole moment $(\Delta\mu)$ relative to that of clean Au(111), the dipole moment of a free-standing SAM $(\mu_{SAM})$ and that of the adsorbate-surface bonds $(\mu_{chem} = \Delta\mu - \mu_{SAM})$; all in D.

| molecule | W | $\Delta W$ | $\Delta\mu$ | $\mu_{SAM}$ | $\mu_{chem}$ |
|---|---|---|---|---|---|
| $CH_3S$ | 3.81 | −1.44 | −0.86 | −0.88 | +0.02 |
| $CH_3CH_2S$ | 3.93 | −1.32 | −0.79 | −0.81 | +0.02 |
| $CF_3S$ | 5.97 | +0.72 | +0.43 | +0.44 | −0.01 |
| $CF_3CH_2S$ | 6.27 | +1.02 | +0.61 | +0.53 | +0.08 |